# Radiative lifetime of localized excitons in transition metal dichalcogenides

Sabrine Ayari[1], Adlen Smiri[1], Aida Hichri[1] Sihem Jaziri[1,2]  and Thierry Amand[3]

[1]Faculté des Sciences de Bizerte, Laboratoire de Physique des Matériaux: Structure et Propriétés, Université de Carthage, 7021 Jarzouna, Tunisia

[2]Faculté des Sciences de Tunis, Laboratoire de Physique de la Matiére Condensée, Déepartement de Physique, Universit Tunis el Manar, Campus Universitaire 2092 Tunis, Tunisia

[3]Université de Toulouse, INSA-CNRS-UPS, LPCNO, 31077 Toulouse, France

sabrineayaari8@gmail.com

*Abstract*

Disorder derived from defects or strain in monolayer TMDs can lead to a dramatic change in the physical behavior of the interband excitations, producing inhomogeneous spectral broadening and localization; leading to radiative lifetime increase. In this study, we have modeled the disorder in the surface of the sample through a randomized potential in monolayer WSe2. We show that this model allows us to simulate the spectra of localized exciton states as well as their radiative lifetime. In this context, we give an in depth study of the influence of the disorder potential parameters on the optical properties of these defects through energies, density of states, oscillator strengths, photoluminescence (PL) spectroscopy and radiative lifetime at low temperature (4K). We demonstrate that localized excitons have a longer emission time than free excitons, in the range of tens of picoseconds or more, and we show that it depends strongly on the disorder parameter and dielectric environment. Finally, in order to prove the validity of our model we compare it to available experimental results of the literature.

## Introduction

Due to the confinement into a single layer and reduced dielectric screening the family of TMDs monolayers are characterized by robust, non-hydrogenic excitonic series with binding energies in the order of hundreds of meV [1-13]. They present strong absorption and short population lifetime, in the picosecond range [13-15]. These features make TMDs attractive materials in electronics [16-18], optoelectronics [19-22] , including photodetectors, light-emitting diodes, and, more recently for lasers [23-27], photonic devices [28,29], and valleytronics [30-32]. For instance, the optical properties of monolayer tungstenides such as WSe2 or WS2 are usually dominated by two direct excitonic transitions (e.g. 1.75 and 1.72 eV in WSe2), known as excitons and trions respectively [2, 4, 6]. As a result of growth kinetics and adsorption when exposed to atmosphere, TMDs crystal lattice can contain a number of different structural defects which significantly alter their properties [33]. In this context, several discrepancies between theory and experiment have shown that structural defects in TMD monolayers play significant roles in the electronic and optical properties [2,6,8,33-37]. For example, the authors of the reference [38] investigate the exciton to trion formation dynamics in monolayer TMD using ultrafast, two-color pump-probe spectroscopy with properly chosen spectral and temporal resolutions. They showed that the trion formation time increases by _ 50% as the pump energy is tuned from the high energy to low energy side of the inhomogeneously broadened exciton resonance. This observation suggests the presence of an effective exciton mobility edge, i.e., below (above) some energy threshold, the center-of-mass motion of the excitons is localized (delocalized) in a disorder potential [39]. TMDs monolayers exhibit rich photoluminescence (PL) spectra which not only show a strong excitonic effect, but also ensue a broad defect-activated emission peak series within the optical band gap, which do not appear in perfect (pristine) crystal structure [2,6,40,41]. At low temperature, excitation power dependence measurements proved the existence of multiple emission peaks in monolayer TMDs akes which are located below the neutral and charged exciton peaks [2, 6, 42-48]. Some of the pronounced peaks in PL are commonly attributed to defect-localized excitons [2, 6,42-49]. For example, Wang et al. in the reference [6] recorded at T=4K an anharmonic emission band (fingers- like pattern ) that can be attributed to the localized exciton between 1.65-1.71eV below the free exciton. These localized excitons are excellent single-photon emitters and they prove that spectral stability is excellent, which renders these localized emitters in monolayer WSe2 promising for applications in optoelectronics and quantum information processing [2,43,44,46,48]. Such observations point to a simple fact: structural defects in TMDs cannot simply be ignored. As a matter of fact, the nature and the origin of these lines remains unknown and still under debate. In our study motivated by the work of Wang et al [6], we assume that these peaks originate in the defects that exist in the monolayer of WSe2, which are unintentionally generated during materials synthesis or when transferring a monolayer onto the substrates[33,38]. Similar effects are observed in WS2 [50].

The large exciton and trion binding energies imply that many body interactions play an important role in determining the optical properties of these materials [6,51-54]. For most optoelectronic applications, knowing the radiative lifetimes of elementary excitations and the parameters that affect this latter is of key importance for realizing applications of these novel materials [6,51-56]. The radiative recombination is an unavoidable process in direct gap semiconductor materials that causes loss of free electrons and holes from the transport process [6,52,54,57]. Recent experimental and theoretical studies also demonstrated that the radiative lifetime of the localized exciton is expected to be more sensitive to the sample parameters (substrate material, interface defects, etc.) than the free exciton emission [6,52,58]. It is shown that both surface defects and the nature of the substrate can strongly affect the light emission decay time of the exciton in monolayer TMDs [6,52,58]. In fact, since radiative rates of excitons are limited by momentum conservation requirements, localization can enhance or reduce radiative rates of excitons by broadening the distribution of their center of mass momenta [51,52]. It is important to note that in recent experimental results the distinction between "localized" and "delocalized" is blurred by the finite lifetime [6,51]. By using the time-resolved photoluminescence (TRPL) Wang et al. in the reference [6] proved that under some experimental condition, these localized excitons are characterized by long emission times,compared to the free exciton, in the tens of picosecond range. Another work proved that the lifetime can reach the nanosecond range for strongly localized exciton [51]. It is therefore of fundamental importance to understand the effect of disorder and the environment in order to determine the radiative emission processes of localized exciton. The interpretation of experimental data and optimization of monolayer TMDs for optoelectronic device applications require a theoretical model for prediction of the energy, the oscillator strength of optical transitions and the radiative lifetime of localized exciton .In a previous work, we had studied the effect of the dielectric environment surrounding the ML-WS2 on the fundamental properties of exciton and trion (size, binding energy, oscillator strength) via Keldysh potential and we had given a quick discussion on the disorder model [59]. In the present work, we go one step further by giving a thorough study of the localization effect in WSe2 monolayers. We calculate the radiative lifetime of the localized exciton in a ML-WSe2 as well as the effect of the characteristic parameters which define the disorder (coherence length, fluctuation amplitude) on the exciton energies, PL spectrum, radiative lifetimes and wave function at low temperature (T=4K). The latter proves to strongly depend on the disorder parameter. We finally compare our results with resent spectroscopy experiments performed on WSe2 in order to prove the validity of our model.

The organization of this article is as follows: in section 2 we introduce our random potential fluctuation model in order to describe the exciton center of mass localization: we discuss the relevant parameter range for the model and , using the Fermi golden rule, we calculate the radiative lifetime of the localized exciton at low temperature in ML-WSe2; in section 3 in order to provide further insight into the behavior and origin of the excitonic localized states in ML-WSe2, we discuss the influence of the coherence length of the random disorder potential on the optical properties at T=4K: on the basis of PL spectrum , emission energies, probability distributions of excitonic states and radiative lifetime in specific disorder realizations.; in the section 4, we investigate the influence of the disorder potential fluctuation amplitude. The dielectric environment effect on the radiative lifetime of the localized exciton at low temperature in ML-WSe2 will be studied in section 5; in section 6 we compare our results with time- resolved photoluminescence (TRPL) spectroscopy experiements available in the literature . Finally in the last section we present our conclusions.

## *2. Formulation of the localized exciton problem*

In this section, we will present the study of the effect of the disorder in three step: In step 1, we calculate the eigenstates and eigenvalues of a localized exciton by solving the effective Schrodinger equation; In step 2, in order to understand the framework of the disorder potential we will give a theoretical description of the influence of the disorder parameters (coherence length, fluctuation amplitude) on the optical properties. In step 3, by using the Fermi golden rules and the oscillator strength evaluated in step 1 we will calculate the radiative lifetime of the localized exciton.

### *2.1 Description of the Hamiltonien of the localized exciton via disorder potential*

To model the localization of the exciton, we choose the case of a spatially random disorder potential that binds the center of mass of the electron-hole pair and create inhomogeneous multiple emission peaks on the low energy side of the emission spectra, with long radiative lifetime compared to the free exciton one. To achieve this, we describe the disorder potential by a Gaussian random field for electrons and hole $V_{e,(h)}(\boldsymbol{\rho}_{e,(h)})$, created from a superposition of N random plane waves with random direction $\theta_i$, random phase $\phi_i$, and a wavelength L which will be shown to correspond to the disorder potential correlation length:

$$V_{e,(h)}(\boldsymbol{\rho}_{e,(h)}) = \frac{V_{0_{e,(h)}}}{\sqrt{2N}} \sum_{i=1}^{N} \Re e\{e^{i(K.\rho_{e,(h)}+\phi_i)}\} \qquad (1)$$

with $\boldsymbol{K} = \left(\frac{2\pi}{L}\cos\theta_i, \frac{2\pi}{L}\sin\theta_i\right)$, $\boldsymbol{\rho}_e$ and $\boldsymbol{\rho}_h$ are the in-plane position vectors for the electron and the hole respectively and $V_{0_{e,(h)}}$ is the fluctuation amplitude for the electrons and hole respectively. This potential is characterized by zero-spatial average $\overline{V(\boldsymbol{\rho}_{e,(h)})} = 0$ and a constant variance $\sigma_V^2 = \overline{V(\boldsymbol{\rho}_{e,(h)})^2}$. See appendix A for mathematical details about disorder potential. Apart from the random confinement disorder potential for electron and holes $V_{e,(h)}(\boldsymbol{\rho}_{e,(h)})$, the electron-hole coulomb interaction is treated here using the Rytova-Keldysh potential $V_{Ky}(\boldsymbol{\rho}_e - \boldsymbol{\rho}_h)$ according to the widely accepted approach [59,60,61-63], in order to take properly into account the dielectric screening of the surrounding

[10,11,59,60]. According to the effective mass theory and in the center-of-mass frame, the Hamiltonian can be written in the form:

$$H = \frac{-\hbar^2 \nabla_\rho^2}{2\mu} + V_{Ky}(\boldsymbol{\rho}) - \frac{\hbar^2 \nabla_R^2}{2M_X} + V_e\left(\boldsymbol{R} - \frac{m_e}{M_X}\boldsymbol{\rho}\right) + V_h\left(\boldsymbol{R} + \frac{m_h}{M_X}\boldsymbol{\rho}\right) \qquad (2)$$

Here $\boldsymbol{R} = \frac{m_e \boldsymbol{\rho}_e + m_h \boldsymbol{\rho}_h}{m_e + m_h}$ and $\boldsymbol{\rho} = \boldsymbol{\rho}_e - \boldsymbol{\rho}_h$ are the position vector of the exciton center of mass and relative distance of the electron from the hole respectively, $M_X = m_e + m_h$ is the exciton mass, $\mu = \frac{m_e m_h}{m_e + m_h}$ is the reduced effective mass, and $\nabla_\rho$, $\nabla_R$ are the gradient operators acting on the relative and center of mass coordinates respectively.

We shall assume as working assumptions that i) the random potential spatial correlation length is much larger than the free exciton Bohr radius $a_b$ (L>> $a_b$), so that the exciton internal motion is not perturbated by localization [64,65,66]. ii) The perturbation introduced by disorder is not sufficient to produce a significant coupling between the $\widetilde{1s}$ exciton state and higher excited states of the electron hole motion [64,65,66]. Hence only the lowest bound state $\widetilde{1s}$ at the fundamental sublevel transition will be considered. The expectation expression of the effective random disorder potential $V_{e,(h)}(\boldsymbol{\rho}_{e,(h)})$ under this assumption is given by the expression:

$$\hat{V}_{\widetilde{1s}}(\boldsymbol{R}) = \left\langle \chi_{\widetilde{1s}} \left| V_e\left(\boldsymbol{R} - \frac{m_e}{M_X}\boldsymbol{\rho}\right) + V_h\left(\boldsymbol{R} + \frac{m_h}{M_X}\boldsymbol{\rho}\right) \right| \chi_{\widetilde{1s}} \right\rangle = \frac{2\pi}{\sqrt{2N}} \sum_{i=1}^{N} \Re e\{e^{i(\boldsymbol{K}\cdot\boldsymbol{R}+\phi_i)}\} \times \mathfrak{T}(\widetilde{1s}, \widetilde{1s}, K) \quad (3)$$

With $\mathfrak{T}(\widetilde{1s}, \widetilde{1s}, K) =$

$$\sum_{n,l} C(n,l) \sum_{n',l'} C(n',l') i^{l'-l} \int \varphi_n(\rho)\varphi_{n'}(\rho)\left(V_{0e} J_{l'-l}\left(\frac{m_e}{M_X}K\rho\right) + V_{0h} J_{l'-l}\left(\frac{m_h}{M_X}K\rho\right)\right)\rho d\rho,$$

Here $|\boldsymbol{K}| = \frac{2\pi}{L}$, $\chi_{\widetilde{1s}}(\rho,\theta) = \sum_{n,l} C(n,l)\varphi_{n,l}(\rho,\theta)$ is expanded in terms of 2D-hydrogenic state $\varphi_{n,l}(\rho,\theta)$ (see ref [59,67] for more details of the calculation of the wave function $\chi_{\widetilde{1s}}(\rho,\theta)$) and $J_{l'-l}(K\rho)$ is the Bessel function of the first kind. By using the above assumption and for typical values of $V_{0h} = V_{0e} = V_0$, the effective random disorder potential, can be written as :

$$\hat{V}_{\widetilde{1s}}(\boldsymbol{R}) = \frac{V_0}{\sqrt{2N}} \sum_{j=1}^{N} \Re e\{e^{i(KR+\phi_j)}\} \times \mathfrak{T}(\widetilde{1s}, \widetilde{1s}, K) \qquad (4)$$

With $\mathfrak{T}(\widetilde{1s}, \widetilde{1s}, K) = \left(\left[1 + \left(\frac{2\pi m_h a^*_b}{4M_X L}\right)^2\right]^{-\frac{3}{2}} + \left[1 + \left(\frac{2\pi m_e a^*_b}{4M_X L}\right)^2\right]^{-\frac{3}{2}}\right)$. Here $a^*_b$ is the effective Bohr radius in which we have taken into account the contribution of all the hydrogenic states. In our case, for values of L much larger than $a^*_b$ (L>>$a^*_b$), $\mathfrak{I}(\widetilde{1s}, \widetilde{1s}, K)$ tend to 2. Hence, we can conclude that in this limit the disorder affects only the center of mass motion and $\hat{V}_{\widetilde{1s}}(R)$ is given by:

$$\hat{V}_{\widetilde{1s}}(R) = V(R) = \sqrt{\frac{2}{N}} V_0 \sum_{i=1}^{N} \cos(\frac{2\pi}{L} X\cos\theta_i + \frac{2\pi}{L} Y\sin\theta_i + \phi_i) \quad (5)$$

Where (**X**,**Y**) are the in plane **R** coordinates. In this case, we can separate the exciton center of mass and relative motion. The Hamiltonian of relative motion $H_{rel}$ is largely studied in previous work [59,67]. The Hamiltonian which describes the localization of the center of mass motion by a disorder potential can be written as follows:

$$H_{cm} = -\frac{\hbar^2 \nabla_R^2}{2M_X} + V(R) \quad (6)$$

Using the method of numerical diagonalization, we calculate the eigenvalues $\tilde{E}_{(j)}$ and the eigenfunctions $\zeta_{(j)}(R)$ solution of this Hamiltonien. In order, to solve the eigenvalue equation, we use a wave function expansion technique; so $\zeta_{(j)}(R)$ can be factorized into: $\zeta_j(R) = \sum_{n_x,n_y} D(nx,ny)\xi_{n_x,n_y}(X,Y)$ where $\xi_{n_x,n_y}(X,Y)$ are the wave functions of a 2D harmonic oscillator, which are constructed using the Hermite polynomials $\mathcal{H}_{nx,ny}(X,Y)$ and are given by: $\xi_{nx,ny}(X,Y) = \frac{1}{\sqrt{n_x!n_y!}} \frac{1}{\sqrt{2^{(n_x+n_y)}}} \sqrt{\frac{1}{\pi R_{cm}^2}} \mathcal{H}_{n_x}\left(\frac{X}{R_{cm}}\right) \mathcal{H}_{n_y}\left(\frac{Y}{R_{cm}}\right) e^{\frac{-X^2}{2R_{cm}^2}} e^{\frac{-Y^2}{2R_{cm}^2}}$.

In this notation, $n_x, n_y$ are the quantum numbers, $R_{cm} = \sqrt{\frac{\hbar}{M_X \omega_{cm}}}$ is the localization radius in the harmonic potential description, $\omega_{cm}$ is the frequency of a 2D harmonic oscillator. The coefficients $D(nx,ny)$ are obtained by solving the matrix problem with eigenfunctions $\zeta_{(j)}(R)$ and eigenvalues $\tilde{E}_{(j)}$. Here, the number (j) refers to the dominant contribution of the coefficients $D(nx,ny)$. In the numerical calculations, the number of basis functions $\xi_{nx,ny}(X,Y)$ should be finite. Usually the basis functions are chosen provided that they are the lowest energy states of the Hamiltonian. This approach leads to a good convergence of the eigenvalues as function of the number of basis states. Therefore, the solutions of the resulting Schrdinger equation of the system satisfy the eigenequation $(H_{cm} + H_{rel})Y_{(\widetilde{1s},j)}(\mathbf{R}, \boldsymbol{\rho}) = E_{(\widetilde{1s},j)} Y_{(\widetilde{1s},j)}(\mathbf{R}, \boldsymbol{\rho})$ are given by: $Y_{(\widetilde{1s},j)}(\mathbf{R}, \boldsymbol{\rho}) = \zeta_{(j)}(R) \chi_{\widetilde{1s}}(\boldsymbol{\rho})$ and $E_{(\widetilde{1s},j)} = E_g + \tilde{E}_{(j)} + E_{\widetilde{1s}}$. We note that the choice of the basis which are Hermite polynomials multiplied by Gaussian functions is based to the fact that we treat only the case where the excitonic state are located in the patential well, i.e. the energy levels are small compared to the fluctuation potential, which can be verified only if ($V_0 >$

$\hbar\omega_{cm}$), with $\hbar\omega_{cm}$ is the confinement energy. In the contrary case ($V_0 < \hbar\omega_{cm}$) the center of mass wave function of the localized exciton cannot be described in the used basis (the choice of the basis function will be discussed in detail in the next paragraph and in the appendix B).

In the presence of random disorder potential, the momentum of the center-of-mass motion is no longer a good quantum number as was assumed in the previous model related to the free exciton [38,49,59]. However, the exciton in monolayer TMDs behaves as a massive particle subject to a disordered potential, leading to spatially localized eigenstates of the center of mass motion [44,38,65,66] and therefore to the longer exciton lifetimes emission [6,51].

### 2.2. Relevant parameter range for the model

We discuss now the impact of the characteristics parameters of the disorder like correlation lengths and amplitude fluctuations potential on the outcome of the simulations. This provides us with further insights into the structure of the single exciton states and their optical properties. The center of mass exciton motion defined by the Hamiltonian Eq (6) and the random disorder potential Eq. (5) depends on the two reduced independent parameters, $\frac{L}{R_{cm}}$ and $\frac{V_0}{\hbar\omega_{cm}}$. In fact, since $R_{cm}$ constitutes the natural length unit for the Schrodinger problem and $\hbar\omega_{cm}$ is the corresponding unit of energy, we can drive the scaling properties of the matrix element of the center of mass motion by using these two units. Therefore, the matrix element of $\widetilde{H_{cm}} = H_{cm}/\hbar\omega_{cm}$ is rewritten as follows:

$$\left\langle \xi_{n_x,n_y} \left| \widetilde{H_{cm}} \right| \xi_{n'_x,n'_y} \right\rangle = \langle \tilde{E}_c \rangle_{n_x,n_y,n'_x,n'_y} + \frac{V_0}{\hbar\omega_{cm}} \sum_{i=1}^{N} \mathcal{F}(n_x, n_y, n'_x, n'_y, \alpha_i, \beta_i) \quad (7)$$

The kinetic term in the equation is $\langle \tilde{E}_c \rangle_{n_x,n_y,n'_x,n'_y}$, The discussion will focus on the second term in the equation by evaluating the potential energy $\langle \tilde{E}_p \rangle_{n_x,n_y,n'_x,n'_y}$. These two contributions are described in more details in Appendix B. Here $\alpha_i = 2\pi \frac{R_{cm}}{L} \cos\theta_i$, $\beta_i = 2\pi \frac{R_{cm}}{L} \sin\theta_i$. The tilde denotes scaled quantities with $R_{cm}$ as length unit and $\hbar\omega_{cm}$ as energy unit

In this paper we revisit our theoretical model given in the ref .[67] and correlate the different characteristic parameters, where $\hbar\omega_{cm}$ gives the disorder potential amplitude $V_0$ and the localization radius in the harmonic potential description $R_{cm}$. This later gives the correlation length L .With this consideration we will discuss two situations, the first one is relative to the amplitude fluctuation and the second concerns the effect of correlation length of the disorder potential :

### a) the effect of the amplitude fluctuation on the localized exciton properties:

For a fixed values of $\hbar\omega_{cm}$ and correlation length L, the parameter $\frac{V_0}{\hbar\omega_{cm}}$ has been adjusted by varying the disorder potential amplitude $V_0$. In fact, two limiting cases are of importance $\frac{V_0}{\hbar\omega_{cm}} > 1$ and $\frac{V_0}{\hbar\omega_{cm}} < 1$. In fact, as shown in paragraph 1 of Appendix B, the choice of the harmonic oscillator basis is only valid if the energy levels found are lower than the potential fluctuations so we may admit that the contribution of the solutions $b_{2v}(X)$ is not important, and that the solutions found are relatively accurate. Hence, in order to respect this condition, it is necessary to restrict in our calculation to the case when $V_0 > \hbar\omega_{cm}$. In this case the center of mass eigenenergies satisfy $\tilde{E}_{(j)} < 0$ which means that excitons are localized in deep potential traps associated with certain types of impurities or edge states at the sample boundaries. In the opposite case ~$V_0 < \hbar\omega_{cm}$ we can obtain positive values of the center of mass energies. However, the states obtained then are unphysical since they cannot be described in principle in the frame of our model using only one type of basis functions (see appendix B). This case would lead in principle to unbound excitons states, which description is out the scope of the present paper.

### b) the effect of the correlation length on the localized exciton properties:

For a fixed value of $V_0$, and $R_{cm}$, the parameter $\frac{L}{R_{cm}}$ has been adjusted by varying the correlation length L. From the equation (7) we notice that there are also two limiting cases. In fact when:

i) $\frac{L}{R_{cm}} \gg 1$ the potential energy $\langle\tilde{E}_p\rangle_{n_x,n_y,n'_x,n'_y}$ will increases with the decrease of the ratio $\frac{R_{cm}}{L}$ i,e with the increasing of L. This can lead to the strong disorder and large correlation length L so to the stronger confinement excitonic energies. Therefore, the shift between the neutral exciton and the first state of localized exciton ($\Delta E = E_{X_0} - E_{LX_1}$) increases with the decreasing of the ratio $\frac{R_{cm}}{L}$. In fact, this effect introduces an additional confinement in all directions like in the case of a quantum dot potential or exciton in narrow semiconductor quantum wells grown by molecular beam epitaxial [44, 64-66].

ii) $\frac{L}{R_{cm}} \ll 1$ the potential energy $\langle\tilde{E}_p\rangle_{n_x,n_y,n'_x,n'_y}$ will deacreases with the decrease of the ratio $\frac{R_{cm}}{L}$, i,e with the increasing of L. therefore the shift ($\Delta E = E_{X_0} - E_{LX_1}$) decreases with the decreasing of the ratio $\frac{R_{cm}}{L}$. This can lead to the weak disorder and short correlation length L.

In conclusion, the disorder effect on the excitonic properties can be controlled by these two dimensionless ratio $\frac{R_{cm}}{L}$ and $\frac{V_0}{\hbar\omega_{cm}}$. In this article we will be only interested in the case where $\frac{V_0}{\hbar\omega_{cm}} \geq 1$ and $\frac{L}{R_{cm}} \gg 1$, where we have strong disorder (confinement) and large correlation length. Indeed, we assume that disorder correlation length L is larger than the exciton Bohr radius. We note that at this stage we do not make any further assumptions for the disordered potential $V(X,Y)$. We point out, however, that the optical properties of the localized exciton are characterized by typical values of the correlation length L, and potential amplitude fluctuation $V_0$ that will be discussed in the next section. These disorder parameters are currently of significant interest, particularly in the estimation of the radiative lifetimes of localized excitons in monolayer TMD semiconductors.

### 2.3 Calculation of radiative lifetime

Once the exciton wave functions have been obtained by diagonalizing the Hamiltonian of the localized exciton, using the dipole matrix elements relevant to interband optical transitions we can calculate the radiative lifetime of localized excitons in monolayers TMDs under the approximations stated above. Since we work at low temperature and under low-density excitation.

i) We do not take into account the thermalization processes. In fact, in our work we do not include scattering by optical phonons, assuming the excitonic temperature to be much lower than the optical phonon energy (30 meV for WSe2) [51,54,68] . ii) Besides, we assume that exciton-exciton scattering and annihilation is inefficient [51,54,68]. Therefore, it can be ignored when calculating the exciton radiative lifetimes. These assumptions mean that our calculation is not valid at high temperatures and at high exciton density where a competition between the radiative and the nonradiative decay takes place. In other hand, TMDs can exhibit bright, dark and Z-modes excitons. Dark exciton states do not couple directly to radiation and thus they will not be considered here [51,54]. So we assume an initial state consisting of a bright exciton with in-plane momentum **K** . The localized Z-modes excitons [15], which could couple to in-plane propagating optical modes with polarization perpendicular to the TMD plane, could be calculated. However, we will disregard them here, since they emit at lower energies and their oscillator strength is quite weak with respect to bright excitons [15, 69].

We start by giving a theoretical description of the spontaneous emission due to the interaction of the localized excitons with the continuum of vacuum photon modes. Under low-density excitation, in the Coulomb gauge, the light-matter interaction is described by the Hamiltonian $\widehat{H}_{int} = \frac{e}{m_0 c}\mathbf{p}.\mathbf{A}(\mathbf{r},t)$, where e is the elementary charge, **p** is the electron momentum operator, c is the light velocity and **A(r**, t) is the vector potential operator in the second quantification:

$$\mathbf{A}(\mathbf{r},t) = \sum_{q,\lambda} \sqrt{\frac{2c\pi\hbar}{qVn_0}} \left\{ \boldsymbol{\epsilon}_q^{(\lambda)} a_q^{(\lambda)} e^{i(\mathbf{q}\mathbf{r}-\omega_q t)} + \boldsymbol{\epsilon}_q^{(\lambda)} a_q^{+(\lambda)} e^{-i(\mathbf{q}\mathbf{r}-\omega_q t)} \right\} \quad (8)$$

Here, the sum is extended to all plane-wave eigenmodes $(\mathbf{q},\lambda)$ within a normalization volume $V = L_z S$ (S is the TMD surface), $\boldsymbol{\epsilon}_q^{(\lambda)}$ is a unit vector in the direction of the polarization $\lambda$. The operator

$a_q^{+(\lambda)}(a_q^{(\lambda)})$ creates (annihilates) a photon with wave vector **q** and polarization $\lambda$. For a given eigenmode, the optical angular frequency is defined by $\omega_q = \frac{c}{n_0}|q|$, where $n_0 = \sqrt{\frac{\varepsilon_1+\varepsilon_2}{2}}$ is the effective optical refraction index of the crystal environment.

The light-matter coupling can be evaluated in the basis $\{|......,n_{q,\lambda}.........\rangle \otimes |\Phi_{(j)}^{\nu}\rangle\}$. Here, $\{|......,n_{q,\lambda}.........\rangle\}$ are the electromagnetic field states in Fock representation. The solution of Schrodinger equation for a localized exciton in a solid,

$$\Phi_{(j)}^{\nu}(\boldsymbol{R},\boldsymbol{\rho}) = \Upsilon_{(j)}(\boldsymbol{R},\boldsymbol{\rho})\Psi^{\gamma_X}(\boldsymbol{r}_e,\boldsymbol{r}_h) \quad (9)$$

$\Upsilon_{(j)}(\boldsymbol{R},\boldsymbol{\rho})$ is the envelope function described in section 2, $\boldsymbol{r}_e = (\boldsymbol{\rho}_e, z_e)$, $\boldsymbol{r}_h = (\boldsymbol{\rho}_e, z_h)$ are the electron and hole spatial coordinates respectively. The function $\Psi^{\gamma_X}(\mathbf{r}_e,\mathbf{r}_h) = \sum_{\alpha_c,\alpha_h} C_{c,\alpha_c;h,\alpha_h}^{\gamma_c,\gamma_h} U_{c,\alpha_c}^{\gamma_c}(\mathbf{r}_e) U_{h,\alpha_h}^{\gamma_h}(\mathbf{r}_h)$ represents the combination of electron and hole Bloch amplitude products transforming according to the representation $\gamma_X$ of the D$_{3h}$ symmetry group [69]. Here $\left|U_{c,\alpha_c}^{\gamma_c}\right\rangle \left(\left|U_{h,\alpha_h}^{\gamma_h}\right\rangle\right)$ is the conduction (hole) Bloch amplitude transforming along the representation $\gamma_c$ ($\gamma_h$) of D$_{3h}$ symmetry group, and $\alpha_{c(h)} = (\tau_{c(h)}, s_{c(h)})$ characterize the single particle valley ($\tau_c = \pm 1$) and effective spin ($s_c = \pm 1/2$) indexes. The hole state is related to the valence electron state by $\left|U_{h,\alpha_h}^{\gamma_h}\right\rangle = \hat{K}\left|U_{v,\alpha_v}^{\gamma_v}\right\rangle$, $\hat{K}$ being the time-reversal operator [70], and the coefficients $C_{c,\alpha_c;h,\alpha_h}^{\gamma_c,\gamma_h}$ are deduced from [15,71]. According to the time dependent perturbation theory, the radiative lifetime is calculated using Fermi's Golden Rule $\tau_{rad}^{-1} = \frac{2\pi}{\hbar}|\langle i|H_{int}|f\rangle|^2 \delta(E_i - E_f)$. The initial state consists of an excitonic state without a photon $|i\rangle = |\Phi_{(j)}^{\nu}\rangle \otimes |0_{q,\lambda}\rangle$ while the final state consists of the crystal ground state $|\emptyset\rangle$ with one photon $|f\rangle = |\emptyset\rangle \otimes |1_{q,\lambda}\rangle$. Therefore, the probability of spontaneous emission of a photon in the mode $(q,\lambda)$ per time unit is then given by:

$$P_{(j),q,\lambda} = \frac{(2\pi)^2 e^2}{n_0 m_0^2 cV} \frac{1}{q}|M_{(j),q,\lambda}|^2 \delta(\hbar\omega_{(j)}^X - \hbar\omega_q) \quad (10)$$

Here, the $\delta$ function ensures the energy conservation between the localized exciton ($E_{\widetilde{1s},(j)} = \hbar\omega_{(j)}^X$) and the photon ($\hbar\omega_q$). We note that for each direction of propagation **q** of emitted photon, there are two independent linear polarizations $\lambda$. The factor $M_{(j),\mathbf{q},\lambda} = \left\langle \Phi_{(j)}^{\nu}\left|\hat{H}_{\mathbf{q}}^{(\lambda)}\right|\emptyset\right\rangle$ is the optical matrix element characterizing the transition from the 2D crystal ground state to the exciton state $\left|\Phi_{(j)}^{\nu}\right\rangle$. For direct excitons, it can be presented under the form:

$$M_{(j),\mathbf{q},\lambda} = \bar{\zeta}_j(\mathbf{K}_X = 0)\chi_{\widetilde{1s}}(\boldsymbol{\rho} = 0)\sum_{\alpha_c,\alpha_h} C_{c,\alpha_c;h,\alpha_h}^{\gamma_c,\gamma_h} \left\langle U_{c,\alpha_c}^{\gamma_c}\left|\boldsymbol{\varepsilon}_{\mathbf{q}}^{(\lambda)}\cdot\hat{\mathbf{p}}\right|U_{v,\alpha_v}^{\gamma_v}\right\rangle \quad (11)$$

where $\bar{\zeta}_j(K_X = 0) = \iint \zeta_j(R)d^2R$ represents the Fourier transform of the centre of mass wave function taken at $K_X = 0$. We have used the fact that the Bloch functions are orthogonal, that the envelope functions are slowly varying on the scale of the lattice parameter, and that the wave vector **q** is negligible compared to the size of the Brillouin zone. By summing $P_{q,\lambda}$ over all the photon modes $(q, \lambda)$, $P_{(j)} = \sum_q P_{q,\lambda}$, we obtain,

$$P_{(j)} = \frac{e^2}{2\pi \hbar c^3 m_0^2} n_0 \omega_j^X \left| \sum_{n_x,n_y} D(nx, ny) \int \xi_{nx,ny}(R) d^2R \right|^2 |\chi_{\widetilde{1s}}(\rho = 0)|^2 I \quad (12)$$

Here, we have used integration over a Dirac delta function in wave vector $\int q\,dq\,\delta(\hbar\omega_{(j)}^X - \hbar\omega_q) = \frac{n_0^2 \omega_{(j)}^X}{\hbar c^2}$. In the above equation, the factor $I = \sum_{\lambda=1,2} \int_0^\pi \int_0^{2\pi} d\theta\,d\varphi \left|\left\langle U_{c,\alpha_c}^{\gamma_c} \left| \varepsilon_{\mathbf{q}}^{(\lambda)}(\theta,\phi).\hat{\mathbf{p}} \right| U_{v,\alpha_v}^{\gamma_v} \right\rangle\right|^2$ represents the angular integral over all the photon modes (including their polarization) that can be spontaneously emitted. The matrix element $\left\langle U_{c,\alpha_c}^{\gamma_c} \left| \varepsilon_{\mathbf{q}}^{(\lambda)}.\hat{\mathbf{p}} \right| U_{v,\alpha_v}^{\gamma_v} \right\rangle = \varepsilon_{\mathbf{q}}^{(\lambda)}.\left\langle U_{c,\alpha_c}^{\gamma_c} \left| \hat{\mathbf{p}} \right| U_{v,\alpha_v}^{\gamma_v} \right\rangle$ appearing in the integral $I$ depends on the nature of Bloch functions and the photon polarization $\varepsilon_{\mathbf{q}}^{(\lambda)}$, and the coupling term $\left\langle U_{c,\alpha_c}^{\gamma_c} \left| \hat{\mathbf{p}} \right| U_{v,\alpha_v}^{\gamma_v} \right\rangle$ can be found in Koster tables within a constant coefficient [71], yielding the exciton chiral selection rules. Taking into account the emission of bright excitons only [15], we obtain $\left|\Psi_{\pm 1}^{\Gamma_6}\right\rangle = \left|U_{c,\pm 1,\pm 1/2}^{\Gamma_9}\right\rangle \left|U_{h,\mp 1,\mp 1/2}^{\Gamma_7}\right\rangle$, which belongs to $\Gamma_6$ representation of $D_{3h}$. The only non-zero elements of the valence conduction coupling term are $\left\langle U_{c,\pm 1,\pm 1/2}^{\Gamma_9} \left| \hat{p}_{\pm} \right| U_{v,\pm 1,\pm 1/2}^{\Gamma_7} \right\rangle = \pm \Pi_\perp$ for circularly polarized light $\sigma^\pm$ propagating along the normal to the sample ($\hat{p}_\pm = (\hat{p}_x \pm i\hat{p}_y)/\sqrt{2}$), so that only optical modes with in-plane polarization components couple to these excitons.

The quantity $\Pi_\perp$ can be approximately evaluated by using the **kp** two band model [57,72], In that case, it is given by $\Pi_\perp = \sqrt{\frac{m_0 E_p}{2}}$ where $E_p = \frac{m_0 E_g}{m_e^*}$ is homogeneous to Kane factor. The integral I is given by $I \sim \frac{16\pi}{3} |\Pi_\perp|^2$. Finally, we obtain the spontaneous emission rate,

$$P_{(j)} = \frac{4e^2}{3\hbar m_0 c^3} n_0 \omega_j^X E_p \left| \sum_{n_x,n_y} D(nx, ny) \int \xi_{nx,ny}(R) d^2R \right| |\chi_{\widetilde{1s}}(\rho = 0)|^2 \quad (13)$$

Using the above relation, we can now calculate the radiative lifetime of localized excitons $\tau_{(j)}^{rad} = \frac{1}{P_{(j)}}$. Having described in this section the theoretical model, we turn now to individual excitonic states in specific disorder realizations. In fact, according to the experimental considerations we can find several scenarios of disorder. In our study by the variation of appropriates parameters L and $V_0$, we realize different case of disorder: weak, mean or strong disorder. We start the discussion with the effect of the correlation length L, later we study the effect of the potential amplitude fluctuations $V_0$.

### 3. Impact of the correlation length on the exciton states in single disorder realizations

To investigate the effect of L on the excitonic properties we carry out several simulations for different values of L. In the following, three realizations I, II and III of disorder potential are discussed and compared, having a fixed $\hbar\omega_{cm}$, $V_0$, N and $R_{cm}$ but various values for L. The simulated potential landscapes $V_{I,II,III}(\mathbf{R})$ are shown in the panel of Fig. 1.(a,b,c). One can observe the localized energies Fig. 1.(d,e,f), probability density function in Fig.2(a,b,c), exciton PL peaks Fig. 3(a,b,c), as well as their corresponding radiative lifetime Fig.3(d,e,f).

#### 3.1 The disorder potential

For WSe$_2$ monolayer deposited on the top of the SiO$_2$ substrate and exposed to the air with $m_e = 0.48 m_0$ $m_e = 0.44 m_0$ [73] and for $N = 1000$, $V_0 = 110 meV$, $R_{cm} = 30$ Å and L= $2.3 R_{cm} = 70$Å (simulation I illustrated in Fig. (1.a)), L=5 $R_{cm}$ =150 Å (simulation II plotted in Fig. (1.b)), L=20 $R_{cm}$ =600 Å (simulation III shown in Fig. (1.c)), the disorder potential is characterized by several inhomogeneous lakes (wells) -and-hills having different trenches of more circular or more elongated and irregular shape [74]. These lakes constitute the potential traps which localize the exciton center-of-mass motion. They are randomly distributed over the whole space extent of the simulation. Indeed, these images prove the random nature of the disorder potential. Interestingly, the distributions, the heights, the depths, the numbers of peaks, wells and the form of potential are different from one simulation to another. This can lead to a variation in the confinement and therefore a variation of the energies, the waves function the oscillator strength and the radiative life time which are the characteristic features of localized excitons. Moreover, we notice from these figures that for a given region of space (here 12.5× 12.5 in the unit of $R_{cm}$) when L increases the numbers of lakes-and-hills decreases. For example, for L ~70Å, figure(.(1.a)) we have several peaks and wells while the disorder potential is characterized by a single large well when L becomes much greater than the center of mass radius $R_{cm}$ ( L >>$R_{cm}$) (figure (1.c)). Also we observe the extension of the wells width with the increase of L. This can lead to a stronger confinement and therefore to a more localized excitonic states.

#### 3.2 Localized excitonic energies.

According to the previous simulations we plot in the panels of Fig. (1.d), Fig. (1.e), Fig. (1.f) the first energy levels of the excitonic states $E_{(\widetilde{1s},j)}$ located in the disorder potential given by equation (5). These solutions are obtained by solving the Hamiltonian $H$ (eq.2). From the realization I and II (figure (1d), (2e)) we found that the localized exciton series in monolayer WSe$_2$ deviates significantly from a 2D harmonic oscillator in two significant ways: i) first, the spacing between the energy levels of the localized exciton are no longer equidistant. For example in the simulation II when L=150 Å the spacing between the second and the first level $\delta E_{(2),(1)}$ is equal 7.65 meV while the spacing between the third and the second level is equal to $\delta E_{(3),(2)} = 18\ meV$. We note that the difference between states is smaller than the fluctuation amplitude $V_0$ ($V_0 > \delta E_{(j+1),(j)}$), ii) In addition, in a two-dimensional harmonic oscillator the states are (2n+1) fold degenerate; however, our calculations show clearly that a splitting of these multiplets can arise, leading to reduced degeneracy. Moreover, the confinement energies and the spacing between the energy levels are smaller compared to the binding energy and the spacing of the energy levels of the delocalized exciton. This decrease in energy-level splitting is caused by a strong spatial-localization of the exciton center of mass in the disorder potential due to the anharmonicity of the disorder potential. Although these two simulations present the same features, the confinement energies and the spacing of the energy levels are completely different. For realization III, we use the same $V_0$= 110meV and $R_{cm}$ = 30Å as the other simulations but an L much greater than the previous simulation (L = 600 Å). From figure (3.b) we can notice that the lowest states of the exciton series is approximately conform to a two-dimensional harmonic oscillator with a quasi $(2(n_x + n_y) + 1)$ fold degeneracy and an approximate equidistance between consecutive energy levels. For that we can use the same notation as a 2D harmonic oscillator. However, this simple description becomes non adequate for the highest excited states, since the local Gaussian disorder potential deviates strongly from the parabolic approximation.

### 3.3 Probability densities of Localized excitons

Following the study of the effect of L on the energy we turn now to discuss its effect on the probability densities of the center-of-mass motion for the three different simulations. According to the profile plotted in fig. (1.(a,b,c)), we display in fig.2 (a,b,c) the square modulus of some exciton center of mass wave functions selected among the lowest-energy eigenstates. In these panels, the excitonic states are labeled by an eigenstate number j, sorted in order of ascending eigenenergies $\tilde{E}_{(j)}$. We begin the discussion with realization I fig.2(a). We have therefore chosen to plot in each panel several states having similar features, according to the discussion below. We notice that for realization I, most of the states below the neutral exciton PL peak are attributed to the local ground states because they are formed by a single lobe . i.e., has a dominant contribution from a local ground solution. These results are expected. Indeed, since our potential is characterized by several wells of the same depths in different regions of space, there is a high probability of having similar states that can be located in a local minimum of disorder potential. The probability densities of the center of mass wave function in real space for the first few states of the realization II are illustrated in figure (2.b). The nodal structure of the wave functions is observed in this plot. The state 1 is the exciton ground state of the considered area and thus also a local ground state. States 2,6,8 show a similar behavior, and should thus be attributed to local ground states because they are formed by a single lobe, i.e., has a dominant contribution from a local ground solution. The states 3 and 4 (with 2 lobes) are the first excited states of state 1 while the states 7 (3 lobes) ,9(4 lobes) are further excited states of state 1. We note that all these states are located in the well situated in the middle of the sample. Similarly, states 5 (2 lobes) is excited states of the state 2. We note that each of these states is localized in a local minimum of the disorder potential. In comparison with the realization I we can say generally that while all the states of the simulation I are characterized only by ground states the most of the states in realization II are well described by ground and excited states. For the simulation III the probability density of states are shown in Figure 2(c). We notice that these densities share similar in-plane nodal structures with the excited states in a two-dimensional harmonic oscillator which have the degeneracies 1,2,3,4.., and therefore enable the eigenstates to be labeled with a principal quantum number $|n_x, n_y\rangle$ Unlike the previous cases, all these states are located in the same place, these results are predicted since our potential is characterized by a single well in the simulated region. The state 1 (with only one lobe) in figure (2.c) is attributed to the local ground state which is non-degenerate: to this state we can associate the wave function $\chi_{0,0}(R)$. The states 2 and 3 with two lobes correspond to the second and third excited states. They are twice degenerate and correspond to the wave function $\chi_{1,0}(R)$ and $\chi_{0,1}(R)$ respectively. These states are already quite deformed due to the non-perfect cylindrical symmetry of the potential trench. These results are important for comparison with spectroscopy experiments of excitonic states performed using high spatial resolution.

### 3.4 Photoluminescence spectroscopy of the localized exciton

At T=4K the PL calculating by using the following expression: $P^{LX(j)}(\omega) \propto |\chi_{\widetilde{1s}}(\rho = 0)|^2 \left|\sum_{n_x,n_y} D(nx, ny) \int \xi_{nx,ny}(X,Y) dX\, dY\right|^2 \mathcal{L}(\hbar\omega - E_{(j)})$ displays several anharmonic peaks which have been tentatively identified as localized excitons, where $\mathcal{L}(\hbar\omega - E_{(j)}) = \frac{\gamma}{\pi(\hbar\omega - E_j)^2 + \pi\gamma^2}$ express the energy conservation given by a Lorentzian. $\gamma$ is the half width at half maximum (an arbitrary broadening parameter has been taken here for all the localized states).In the panels of Fig. 3.(a,b,c) we reproduce the multiple emission peaks observed in monolayer WSe2 for

different value of L(70A°, 150A°, 600A°). We note that the PL features of $X_0$ is attributed to neutral exciton peak from a region without defect (for comparison the free exciton line deduced from a former work for a homogeneous sample is put in the inset of the panel 3.(a) ), whereas the features $LX_j$ correspond to PL from the localized exciton states. In low quality samples with high defect density or non-encapsulated samples, we see comb-like lines in the PL spectra: this spectrum di_ers signi_cantly from the spectral characteristics of pristine WSe2 monolayer where only two peaks were observed at ~ 1,752 and ~ 1,728 eV which are attributed to the neutral and charged exciton respectively [2, 6,42]. In fact, the delocalized exciton emission peak $LX_0$ from the defect site in single layer (fig.3) is significantly quenched in the presence of the localized exciton and is dominated by the red-shifted emission from defects [2,42,44]. The quenching of the delocalized excitons emission is a consequence of the exciton relaxation into dark and localized states [44]. Therefore, disorder acts as exciton traps and reduces the mobility and the discusion length of exciton. Moreover, one immediately notices that the sharp emission peaks originating from localized excitons $LX_j$ are characterized by an anharmonic and inhomogeneous spectrum. Comparing with the delocalized exciton emission in monolayer WSe2, these localized emitters have line-widths almost an order of magnitude sharper than the delocalized excitons [42,43,44]. From Fig. 3(a,b,c) we notice that because of the variation of the correlation length L the PLs spectrum are significantly different from one simulation to another. In order to exhibit the effect of L ,we consider the realization II as a reference and we compare it with the two other simulations. We start the discussion with simulation II (fig 3.b) which correspond to L=150 Å.We notice that the spectrally localized emissions are between the energy range 1.64eV - 1.71eV which is lower in energy from the PL of the neutral exciton peak of WSe2 (1.752 eV) range. It is located about 107 meV below the PL peak of neutral free exciton. We compare these results with the PL spectra illustrated in Fig. 3(a), Fig. 3(c) and associated to the simulation I and III, we notice that the energy shift ($E_{X0} - E_{LX1}$) between the free and the ground localized exciton and the shape of the curves change completely. In the simulation I which correspond to L=70Å figure (3.a) these defects introduce new emission peaks at 40 meV below the free-exciton $X_0$. This narrow spectrum shows a broad emission resonances located around 1.712 and 1.716eV. This can be explained as follows: the exciton in this simulation is less localized compared to the first case, i.e less con_ned while for the realization III plotted in fig 3.c , the defects introduce a large localized exciton peak between the 1.54 eV-1.7 eV. From this plot we notice that the energy shift between the localized exciton and the delocalized exciton increases by 113 meV when L increases from 150 to 600 Å. Therefore the exciton in this simulation (L >> $R_{cm}$) is more localized compared to the first and the second case, ie. is more confinement.

### 3.5 Effect of the correlation length L on the radiative lifetime of the localized exciton

For most optoelectronic applications or possible realization of single photon source [2,44], knowing the radiative lifetimes of elementary excitations is critical [6,51-58]. To explore the effect of the three-dimensional quantum confinement of excitons we now proceed to study the effect of L on the radiative lifetime of the localized exciton for the three different simulations by using the relation $\tau_{(j)}^{rad} = \frac{1}{P_{(j)}}$ and by comparing this result with the radiative lifetime of the free exciton. For light propagating perpendicular to the 2D layer and by assuming conservation of the in-plane wavevector, the radiative lifetime of the free exciton can be calculated using the expression given in the reference [57]. Thus, we find a fast free exciton radiative lifetime on the order of 1 ps, this shorter life time can be explained by the small exciton radius and the large delocalized exciton optical oscillator strength inWSe$_2$ monolayers [51]. However, as shown in figure 3( d,e,f ),for localized excitons we calculate radiative lifetime much longer than the free exciton: it varies from tens of picoseconds (for $L \sim R_{cm}$) to more than a nanosecond, becoming longer for larger values of correlation length L (i,e for $L \gg R_{cm}$) [6,51]. The calculated lifetimes for the chosen values of R$_{cm}$ and V$_0$ are quantitatively different from one simulation to another. As a matter of fact, when L is of the order of Rcm i.e for a narrow well, the radiative lifetime is in the order of ten ps ( $\tau_{(1)}^{rad} = 22.47$ps for L=70Å) while the radiative lifetime can exceed a nanosecond for larger correlation length of the potential (L≫R$_{cm}$ ), or strong disorder. We can conclude from this data that at low temperature the increase of the local well extension leads to radiative lifetime considerably longer than radiative lifetime of the free exciton. We note also that the different value through each simulation is due to a complex potential landscape with localization sites of widely varying depth for excitons.

### 4. Impact of the fluctuation amplitude on the exciton states

To understand more deeply the defects states, we now turn to study the effect of the fluctuations amplitude (V$_0$) on the PL peaks and on the magnitude order of localized exciton radiative lifetime in WSe$_2$. Different types of disorder potentials and their effects on excitonic properties are illustrated in Fig.4 (a,b) for a fixed values of $\frac{L}{R_{cm}} \sim 3$, and various amplitude values V$_0$= $2\hbar\omega_{cm}$, $11\hbar\omega_{cm}$ and $20\hbar\omega_{cm}$ with $\hbar\omega_{cm} = 10 \, meV$. According to the plots in the inset of fig (4.a) we notice that for fixed value of L=90Å the depth of the potential increases with the increasing of V$_0$, leading to the enhancement of the three dimensional quantum confinement effects and consequently more localized excitonic states. This is observed by i) the increase in the shift between the peaks of the localized exciton in a disorder potential and the peak of the free exciton ($E_{X0} - E_{LX1}$). ii) The enhancement of the radiative lifetime of the localized exciton. In fact, it 'is appear that at low temperature, for $V_0$ of the order of $\hbar\omega_{cm}$ ($V_0 \sim \hbar\omega_{cm}$), the radiative lifetime of excitons is very small, in the order of few picoseconds $\tau_1^{rad} = 12\ ps$, only one order of magnitude longer than the free exciton one. [54]. Such short radiative lifetimes are attributed to excitons weakly localized by a shallow potential [38]. Their spectral position is only slightly redshifted compared to the delocalized states ($E_{X0} - E_{LX1} = 13\ meV$ on fig. 4(b)). For a large sample surface this leads to inhomogeneous broadening of the free exciton resonance (with still a homogeneous contribution) and the possible appearance of a density of state tail. Nevertheless, From the figure 4(a,b)) we can notice that, for $V_0 = 200$ meV ,the PL is strongly redshifted comparing to the first case ($V_0 = 20$ meV) and is located at 134 meV below the free exciton emission peak and we obtain a long radiative lifetime $\tau_1^{rad} = 475\ ps$. We conclude that,for $V_0 \gg \hbar\omega_{cm}$ localized excitons exhibit long lifetime in the region of hundred picosecond. These lifetimes are almost two orders of magnitude longer than those of 2D free excitons in pristine monolayer WSe2. This type of strongly localized exciton has been recently investigated in WSe2 through single photon emission experiments [38,43]. In addition, the Figure 4.(a,b)) show also the decreasing of the spacing between the energy levels from 57 meV to 2.68 meV when the fluctuation amplitude decrease from $V_0 = 200$ meV to $20\ meV$ .

## 5. Effect of the dielectric environment on the radiative lifetime

Because excitons in 2D semiconductors necessarily reside near a surface, their fundamental properties (size, binding energy, oscillator strength, radiative lifetime) are expected to be strongly influenced by any additional screening from the dielectric environment surrounding the monolayer [10,11]. In our previous work we studied the effect of the dielectric environment on the binding energy as well as the PL emissions peaks, now we explore the impact of the dielectric environment on the localized exciton radiative lifetime by evaluating the inuence of the encapsulated in hexagonal boron nitride(hbN), suspended and silicon nitride (Si3N4) , SiO2 substrate -supported monolayer on the emission properties of monolayer WSe2 at low temperature and for fixed values of $V_0 = 110$ meV , L=90Å , and $R_{cm}$= 30Å . As shown in table 1, it is clear that the radiative lifetime of the localized exciton is strongly sensitive to the details of the dielectric environment and the choice of substrate material. In fact, an obvious increasing in the value of the radiative life time is observed due to the change of the ambient from encapsulated to suspended monolayer .Compared to the radiative lifetime of LX1 state of a suspended WSe2 monolayer, the lifetime $\tau_1^{rad}$ increases with the dielectric constant of the substrate. It is noteworthy that in the case of a lack of environmental screening (suspended monolayer ) the lifetime of LX1 state is 4 ps, while it rises to 76 ps ps when deposed on SiO2 substrate, and reaches 350 ps when the monolayer are encapsulated in between the hBN flakes. This change in the magnitude order of the radiative life time with the dielectric environment can be explained by the fact that: the reduced spatial overlap between the electron and the hole, because of the increasing of the dielectric screening, decreases the transition matrix element and results in longer radiative lifetimes.

## 6. Comparison with the experience

Understanding the effect of each parameter of the disorder potential, allows us to control our model and reproduce the experiment .Usually, the WSe$_2$ monolayer is deposited on top of a 90 nm SiO$_2$ substrate [6]. In this section, we will calculate the energy positions and the radiative lifetime of neutral excitons localized at a local minimum of the random potential. Our objective is to check the validity of our model by a comparison between the radiative lifetimes calculated in our model and those observed in the reference [6]. As we said previously the fluctuations of the crystal potential are governed by the parameters ($V_0$, L). In order to reproduce the observed energy positions and the intensity ratios of the different localized states, we have found that the following set of parameters N=1000, $\hbar\omega_{cm} = 10 meV$, $V_0$ = 110 meV, L=90 Å (the blue data in Figure 4(a,b)), give results in good qualitative agreement with the experiment of ref. [6]. In this work the time-resolved measurements of spontaneous emission gave the localized exciton decay times. However, our model calculates the radiative ones. In Fact, the experimental decay time is defined as $\frac{1}{\tau^{decay}} = \frac{1}{\tau^{rad}} + \frac{1}{\tau^{nrad}}$, where $\tau^{nrad}$ is the non radiative lifetime due to the scattering effects. In the following, we shall assume that, due to the strong exciton localization and low diffusion rate at low temperature [75], the non radiative recombination rate is negligible as compared to the radiative recombination rate $P_{(j)}^{rad} \gg P_{(j)}^{nrad}$ ,therefore $\tau^{decay} \sim \tau^{rad}$. Using our model, we found that for the peak labeled $LX_2$ located at 1.7eV, the radiative life time calculated here is about $\tau_{(2)}^{rad} = 26\ ps$. The peak $LX_1$ located at 1.678eV albeit with considerably longer characteristic time of $\tau_{(1)}^{rad} = 76\ ps$ (the blue data in Figure 4(a,b)). This is in reasonable agreement with the experimental data $\tau_{(1)}^{exp} = 80 \pm\ ps$ ($LX_1$ line) and $\tau_{(2)}^{exp} = 32 \pm\ ps$ ($LX_2$ line). This means that our set of disorder parameters can reproduce reasonably well the experimentally observed localized excitons decay time.

**Conclusion:**

In summary, localized exciton in monolayer WSe2 are described within Wannier- Mott exciton model in which we have described the structural imperfections which may arise due to lattice structural defects, residual impurities or adatoms commonly introduced during the mechanical exfoliation process via a random disorder potential. We have studied the effect of the characteristic parameters (spatial fluctuations in amplitude, correlation length) of the disorder potential on the excitonic properties. We have theoretically investigated the evolution of multiple PL emission peaks in monolayer WSe2 as well as their associated energies , wave function at low temperature (4K) and radiative lifetime . We have proved that disorder potential arising from defects strongly a_ects the exciton states, which leads to inhomogeneous broadening of the exciton resonance with the associated density of localized state as well as the appearance of more deeply localized states. We have demonstrated that, comparing with the free excitons, excitons trapped in potential wells, present a significant enhancement of their radiative lifetime .We proved that the exciton radiative lifetime strongly depends on the disorder parameters and the dielectric screening . We found that if the exciton is weakly localized i,e. in the case when $L \sim R_{cm}$ or/and $V_0 \sim \hbar\omega$ (narrow and shallow well) the radiative lifetime is in the tens of picosecond range while it is in the range of hundreds of picosecond to few nanosecond when $L \gg R_{cm}$ or/and $V_0 \gg \hbar\omega$  i,e larger and deeper well . Finally, in order to valid our model we compare our results with the experience presented in the reference [6] taken as an example. Although the discussion in this paper focuses on WSe2 monolayers, the analysis and the results presented here are expected to be relevant to all 2D metal dichalcogenides (TMDs).

**Appendix**

### Appendix A: Random potential calculation

The desired potential $V(\mathbf{R})$ is a Gaussian random field with zero-mean and unit variance. In the following, we describe the construction of this potential and we verify it has the suitable properties. To achieve this, the potential function $V(\mathbf{R})$ is created from a superposition of N random plane waves with random direction $\theta_j$, random phase $\phi_j$, and constant wavelength L. Let the phase at (0,0) be $\phi_j$. then the potential is given by:

$$V(\mathbf{R}) = C\sqrt{N} \sum_{j=1}^{N} \Re e\{e^{i(K(\theta_j)\mathbf{R}+\phi_j)}\} = C' V_0 \sum_{j=1}^{N} \cos(K(\theta_j)\mathbf{R} + \phi_j) \quad (A1)$$

With $V_0 > 0$. and $C' = C\sqrt{N}$ , C will be defined latter. The random variables $\theta_j$ and $\phi_j$ distributed on $[0,2\pi]$ and are uncorrelated so we can consider that $j=(k,l)$ with $k_{max} l_{max}=N$ Letting $\mathbf{K}(\theta_j) = \frac{2\pi}{L}\mathbf{u}(\theta_j)$, with $\mathbf{u}(\theta_j) = (\cos(\theta_j), \sin(\theta_j))$, we introduce the dimensionless position $\mathbf{R}' = \frac{2\pi}{L}\mathbf{R}$ and $V'(\mathbf{R}') = V(\mathbf{R})/V_0$ ,the potential can be rewritten as :

$$V'(\mathbf{R}') = \frac{C}{2}\sum_{k=1}^{k_{max}}\sum_{l=1}^{l_{max}} \left(e^{i[\mathbf{u}(\theta_k).\mathbf{R}'+\phi_l]} + e^{-i[\mathbf{u}(\theta_k).\mathbf{R}'+\phi_l]}\right) \quad (A2)$$

If N is sufficiently large then $V'(\mathbf{R}')$ is a large sum of random variables and by the central limit theorem, $V'(\mathbf{R}')$ is Gaussian.

### 1. mean value of the disorder potential $V'(R')$

The mean value of the normailzed potential $V'(R')$ over all random variables is written as:

$$\langle V'(\mathbf{R}')\rangle = \frac{C}{2N}\sum_{k=1}^{k_{max}}\sum_{l=1}^{l_{max}} \left(e^{i[\mathbf{u}(\theta_k).\mathbf{R}'+\phi_l]} + e^{-i[\mathbf{u}(\theta_k).\mathbf{R}'+\phi_l]}\right) \quad (A3)$$

Note that since $\theta, \phi \in [0, 2\pi]$ their probability distribution functions are $f(\theta) = \frac{1}{2\pi}$ and $f(\phi) = \frac{1}{2\pi}$ and

$$\langle V'(\mathbf{R}')\rangle \underset{N\to\infty}{\sim} \frac{C}{2}\int_0^{2\pi} e^{i\mathbf{u}(\theta).\mathbf{R}'}\frac{d\theta}{2\pi}\int_0^{2\pi} e^{i\phi}\frac{d\phi}{2\pi} + C.C.$$

$$\langle V'(\mathbf{R}')\rangle \underset{N\to\infty}{\to} 0 \quad (A4)$$

### 2. variance of the disorder potential $V'(R')$

Since the potential have zero mean $\langle V'(R')\rangle = 0$ then the variance $\sigma_{V'}^2 = \langle V'(R')^2\rangle$ is given by :

$$\langle V'^2(\mathbf{R}')\rangle = \frac{C^2}{4N}\left[\sum_{k=1}^{k_{max}}\sum_{l=1}^{l_{max}}\left(e^{i[\mathbf{u}(\theta_k).\mathbf{R}'+\phi_l]} + e^{-i[\mathbf{u}(\theta_k).\mathbf{R}'+\phi_l]}\right)\right]^2$$

$$= \frac{C^2}{4N}\left\{\sum_{k,l}^{k_{max},l_{max}}\left(2 + e^{2i[\mathbf{u}(\theta_k).\mathbf{R}'+\phi_l]} + e^{-2i[\mathbf{u}(\theta_k).\mathbf{R}'+\phi_l]}\right) + \right.$$

$$\left. + \sum_{(k,l)\neq(k',l')}\left(e^{i[\mathbf{u}(\theta_k).\mathbf{R}'+\phi_l]} + e^{-i[\mathbf{u}(\theta_k).\mathbf{R}'+\phi_l]}\right)\left(e^{i[\mathbf{u}(\theta_{k'}).\mathbf{R}+\phi_{l'}]} + e^{-i[\mathbf{u}(\theta_{k'}).\mathbf{R}+\phi_{l'}]}\right)\right\} \quad (A5)$$

with the same reason as previous one the variance $\sigma_{V'}^2$ is equal to : $\langle V'(R')^2\rangle_{N\to\infty} = \frac{C'}{2}$ and we want $\sigma_{V'}^2 = 1$, we take $C' = \sqrt{2}$ as a normalization constant. We note that $\sigma_{V'}^2$ is independent of the position $\mathbf{R}$. Finally, the random disorder potential is rewriten as:

$$V'(R') = \sqrt{\frac{2}{N}}\sum_{j=1}^{N}\cos(u(\theta_j)\mathbf{R} + \phi_j) = \sqrt{\frac{2}{N}}\sum_{j=1}^{N}\cos(|R'|\cos(\theta_j - \phi_{R'}) + \phi_j) \quad (A6)$$

with $\mathbf{R}' = |\mathbf{R}'|(\cos\varphi_{\mathbf{R}'}, \sin\varphi_{\mathbf{R}'})$ It follows that the expression of the initial random disorder potential is given by :

$$V(R) = \sqrt{\frac{2}{N}} \sum_{j=1}^{N} \cos(\frac{2\pi}{L}|R|\cos(\theta_j - \phi_R) + \phi_j) \quad (A7)$$

As demonstrated, this potential have zero mean $\langle V(\mathbf{R})\rangle = 0$ and constant variance equal to

$$\sigma_V = \langle V^2(\mathbf{R})\rangle^{1/2} = V_0$$

We note that the disorder potential $V(R)$ has no spatial periodicity. Each space wave j has a random period of $|\Delta \mathbf{R}_j| = \dfrac{L}{\cos(\theta_j - \varphi_{\mathbf{R}})}$. Only the waves propagating exactly in the direction of $\mathbf{R}$ are periodic with translations with a multiple of the wavelength L. This behavior appeared clearly in the figure (fig S.1). The distribution, height, depth and the form of the potential are different from one simulation to another.

### 3. Calculation of the correlation Length of the disorder potential $V'(R')$

the corelation length is given by the expression:

$$C(\mathbf{R}_1, \mathbf{R}_2) \equiv \frac{\langle V(\mathbf{R}_1)V(\mathbf{R}_2)\rangle - \langle V(\mathbf{R}_1)\rangle\langle V(\mathbf{R}_2)\rangle}{\langle V^2(\mathbf{R}_1)\rangle^{1/2}\langle V^2(\mathbf{R}_2)\rangle^{1/2}} = \frac{\langle V(\mathbf{R}_1)V(\mathbf{R}_2)\rangle}{\langle V^2(\mathbf{R}_1)\rangle^{1/2}\langle V^2(\mathbf{R}_2)\rangle^{1/2}} \quad (A8)$$

In the previous paragraph we demonstrated that $\langle V'(R_1')\rangle = \langle V'(R_2')\rangle = 0$ and $\langle V'(R_1')^2\rangle = \langle V'(R_2')^2\rangle = 1$ so the spatial correlation function of the normalized potential between $R_1'$ and $R_2'$ points and can be written as:

$$C'(R_1', R_2') = C\left(\frac{2\pi}{L}R_1, \frac{2\pi}{L}R_2\right) = \langle V'(R_1')V'(R_2')\rangle \quad (A9)$$

First we begin by calculating :

$$V'(\mathbf{R'}_1)V'(\mathbf{R'}_2) = \frac{1}{2}\sum_{k,l}^{k_{max},l_{max}}\sum_{k',l'=1}^{k_{max},l_{max}} \left(e^{i[\mathbf{u}(\theta_k).\mathbf{R}_1' + \phi_l]} + e^{-i[\mathbf{u}(\theta_k).\mathbf{R}_1' + \phi_l]}\right)\left(e^{i[\mathbf{u}(\theta_{k'}).\mathbf{R}_2' + \phi_{l'}]} + e^{-i[\mathbf{u}(\theta_{k'}).\mathbf{R}_2' + \phi_{l'}]}\right)$$

$$\sim \frac{1}{2}\sum_{k,l}^{k_{max},l_{max}} \left(e^{i[\mathbf{u}(\theta_k).(\mathbf{R}_1' - \mathbf{R}_2')]} + C.C.\right) + \sum_{k,l}^{k_{max},l_{max}}\sum_{(k',l') \neq (k,l)}^{k_{max},l_{max}} \left(e^{i[\mathbf{u}(\theta_k).\mathbf{R}_1' - \mathbf{u}(\theta_{k'}).\mathbf{R}_2') + \phi_l - \phi_{l'}]} + C.C.\right)$$

$$+\frac{1}{2}\sum_{k,l}^{k_{max},l_{max}}\sum_{k',l'=1}^{k_{max},l_{max}}\left(e^{i[\mathbf{u}(\theta_k)\cdot\mathbf{R_1}'+\mathbf{u}(\theta_{k'})\mathbf{R_2}')+\phi_l+\phi_{l'}]}+C.C.\right)$$ (A10)

In the limit when N is suficiently large the second and the third summation will give a zero contribution ,hence

$$C'(R_1',R_2') \sim \frac{1}{2}\int_0^{2\pi}\frac{d\theta}{2\pi}\frac{d\phi}{2\pi}(e^{i[u(\theta_k)(R_1'-R_2')]}+C.C.]$$

$$\sim \int_0^{2\pi}\frac{d\theta}{2\pi}\cos\left[|\mathbf{R}_1{'}-\mathbf{R}_2{'}|\cos\left(\theta-\varphi_{\mathbf{R}_1{'}-\mathbf{R}_2{'}}\right)\right]$$ (A11)

Here $\mathbf{R}_1{'}-\mathbf{R}_2{'}=|\mathbf{R}_1{'}-\mathbf{R}_2{'}|\left(\cos(\varphi_{\mathbf{R}_1{'}-\mathbf{R}_2{'}}),\sin(\varphi_{\mathbf{R}_1{'}-\mathbf{R}_2{'}})\right)$

The normalized correlation function is equal to :

$$C'(\mathbf{R'}_1,\mathbf{R'}_2) \sim J_0(|\mathbf{R}_1{'}-\mathbf{R}_2{'}|)$$ (A12)

We get finally:

$$C(R_1,R_2) = J_0(\frac{2\pi}{L}(|\mathbf{R_1}-\mathbf{R_2}|))$$ (A13)

From the equation (A 13) we see that we can define a correlation length for the random potential as $\Lambda = L$

## Appendix B: SCALING PROPERTIES

In our theory the choice of the basis is strongly based on the behavior of the potential. In fact , the basis of the harmonic function which are Hermite polynomials multiplied by Gaussian functions $\{\xi_{n_x,n_y}; n_x, n_y \in \mathbb{N}\}$ is complete only in the case where the harmonic potential from which it derives is not superiorly bounded (infinite parabolic well). Hence, all the states are located in the well. In our case the random potential uctuations are bounded (basically, the potential is limited by its typical amplitude fluctuations $\pm\sigma_V = \pm V_0$). In this case the general solution, compatible with the behavior of our potential ,should be written using the two orthogonal Parabolic cylinder functions of first and second kind. These functions are the solutions of the generic second order differential equation $-y''[X] + (\frac{x^2}{4} - v - \frac{1}{2})y[X] = 0$ .The general solution thus writes:

$$y[X] = C_1 \cdot 2^{-v/2}e^{-\frac{X^2}{4}}\mathcal{H}_v\left[,\frac{X}{\sqrt{2}}\right] + C_2 \cdot 2^{\frac{1+v}{2}}e^{\frac{X^2}{4}}\mathcal{H}_{-1-v}\left[\frac{iX}{\sqrt{2}}\right]$$

$$= C_1 \cdot b_{1,\nu}(X) + C_2 \cdot b_{2,\nu}(X) \qquad (B1)$$

where $C_1$ and $C_2$ are arbitrary constants. The first term is identified with the eigenfunctions of the harmonic oscillator and is always normalizable:

$$\xi_n[x] = \frac{\mathcal{H}_n[X] e^{-\frac{X^2}{2}}}{\sqrt{2^n n! \sqrt{\pi}}} \qquad (B2)$$

However, the second function is necessary in the case where there is a truncated harmonic well (as in the usual case of a finite square quantum well) or more generally superiorly bounded (another generic solution of Schrdinger equation should then be found out of the domains where the potential is considered as harmonic). Since we restrict here to the case where the excitons are confined in the local wells of the sample, i.e. the energy levels are small compared to the potential fluctuations, we can neglect the contributions of the second function $b_{2,\nu}(X)$ and restrict our function basis to the first one $b_{1,\nu}(X)$.

After the choice of the basis, we analyse now the situation where the disorder induced by structural defects of the monolayer 2D crystal affects only the center of mass motion. In this case, the matrix elements of the centre of mass hamiltonian are sum of:

the kinetic energy contribution to the matrix elements in the above equation are given by:

$$\left\langle \xi_{n_x,n_y}(X,Y) \left| \frac{P^2}{2M_X} \right| \xi_{n'_x,n'_y}(X,Y) \right\rangle = \frac{\hbar \omega_{cm}}{4} T(n_x, n_y, n'_x, n'_y). \quad (B3)$$

Here:

$$T(n_x, n_y, n'_x, n'_y) = \left[ (2n_x + 1)\delta_{n_x n'_x} - \sqrt{n_x(n_x - 1)}\delta_{n_x, n'_x - 2} - \sqrt{(n_x + 2)(n_x + 1)}\delta_{n_x, n'_x + 2} \right] \delta_{n_y, n'_y} + \left[ (2n_y + 1)\delta_{n_y, n'_y} - \sqrt{n_y(n_y - 1)}\delta_{n_y, n'_y - 2} - \sqrt{(n_y + 2)(n_y + 1)}\delta_{n_y, n'_y + 2} \right] \delta_{n_x, n'_x} \quad (B4)$$

the potential energy contribution to the matrix elements in the above equation are in turn given by:

$$\left\langle \xi_{n_x,n_y}(X,Y) \left| V(X,Y) \right| \xi_{n'_x,n'_y}(X,Y) \right\rangle = \frac{V_0}{R^2_{cm}\pi} \sqrt{\frac{2}{N}} \times \frac{1}{\sqrt{2^{n_x+n_y+n'_x+n'_y}}} \times \frac{1}{\sqrt{n_x! n_y! n'_x! n'_y!}} \times$$

$$\sum_{j=1}^{N} \iint_{-\infty}^{+\infty} \mathcal{H}_{n_x,n'_x}\left(\frac{X}{R_{cm}}\right) \mathcal{H}_{n_y,n'_y}\left(\frac{Y}{R_{cm}}\right) e^{\frac{-X^2}{R^2_{cm}}} e^{\frac{-Y^2}{R^2_{cm}}} \cos\left(\frac{2\pi}{L}\cos\theta_j X + \frac{2\pi}{L}\sin\theta_j Y + \phi_j\right) dX dY \quad (B5)$$

The kinetic contribution in the equation is thus known, therefore the whole problem is to evaluate the matrix element which contains the potential energy. This term contains the impact of the disorder potential in the determining the exciton center of mass (COM) spectral properties. The exciton COM problem defined by the Hamiltonian Eq (2) in the main text and the random disorder potential Eq. (4) depends on 2 independent parameters, the correlation length L, the amplitude fluctuations potential $V_0$ Taking the auxiliary basis $\xi_{n_x,n_y}(X,Y)$ a third parameter appears which is the energy quantum ~ $\hbar\omega_{cm}$; the corresponding localization length is $R_{cm} = \sqrt{\frac{\hbar}{M_X\omega_{cm}}}$ These quantities will be taken as energy and length units in the following. Let us introduce the dimensionless coordinates $X' = \frac{X}{R_{cm}}$, $Y' = \frac{Y}{R_{cm}}$ Dividing the matrix element by $\hbar\omega_{cm}$ and letting $\alpha_j = 2\pi\frac{R_{cm}}{L}\cos\theta_j$, $\beta_j = 2\pi\frac{R_{cm}}{L}\sin\theta_j$ we get

$$\left\langle \xi_{n_x,n_y}(X,Y) \middle| H_{cm} \middle| \xi_{n'_x,n'_y}(X,Y) \right\rangle = \tilde{T}(n_x, n_x, n'_x, n'_y) + \tilde{E}_p(n_x, n_x, n'_x, n'_y)$$

$$= \frac{1}{4} T(n_x, n_x, n'_x, n'_y) + \frac{V_0}{\hbar\omega_{cm}} A_{n_x, n_y, n'_x, n'_y} e^{-\left(\frac{\pi R_{cm}}{L}\right)^2} \sum_{j=1}^{N} \mathcal{F}(n_x, n_y, n'_x, n'_y, \alpha_j, \beta_j) \quad (B6)$$

Here $A_{n_x, n_y, n'_x, n'_y} = \frac{2^{n_x+n_y+\frac{1}{2}} n_x! n_y! (-1)^{n_x+n_y}}{\sqrt{N}} \times \frac{1}{\sqrt{2^{n_x+n_y+n'_x+n'_y}}} \times \frac{1}{\sqrt{n_x! n_y! n'_x! n'_y!}}$ (B7)

$$\mathcal{F}_{n_x, n_y, n'_x, n'_y}(\alpha_j, \beta_j) = \cos\phi_j \, \mathcal{M}_{n_x,n_y,n'_x,n'_y}(\alpha_j, \beta_j) - \sin\phi_j \, \aleph_{n_x,n_y,n'_x,n'_y}(\alpha_j, \beta_j) \quad (B8)$$

with

$$\mathcal{M}_{n_x,n_y,n'_x,n'_y}(\alpha_j, \beta_j) = \alpha^{2|m_x|} \beta^{2|m_y|} L_{n_x}^{2|m_x|}\left(\frac{\alpha^2}{2}\right) L_{n_y}^{2|m_y|}\left(\frac{\beta^2}{2}\right) \delta_{n'_x,n_x+2m_x} \delta_{n'_y,n_y+2m_y} -$$

$$\alpha^{2|m_x|+1} \beta^{2|m_y|+1} L_{n_x}^{2|m_x|+1}\left(\frac{\alpha^2}{2}\right) L_{n_y}^{2|m_y|+1}\left(\frac{\beta^2}{2}\right) \delta_{n'_x,n_x+2m_x+1} \delta_{n'_y,n_y+2m_y+1} \quad (B9)$$

$$\aleph_{n_x,n_y,n'_x,n'_y}(\alpha_j, \beta_j) = \alpha^{2|m_x|+1} \beta^{2|m_y|} L_{n_x}^{2|m_x|+1}\left(\frac{\alpha^2}{2}\right) L_{n_y}^{2|m_y|}\left(\frac{\beta^2}{2}\right) \delta_{n'_x,n_x+2m_x+1} \delta_{n'_y,n_y+2m_y} +$$

$$\alpha^{2|m_x|} \beta^{2|m_y|+1} L_{n_x}^{2|m_x|}\left(\frac{\alpha^2}{2}\right) L_{n_y}^{2|m_y|+1}\left(\frac{\beta^2}{2}\right) \delta_{n'_x,n_x+2m_x} \delta_{n'_y,n_y+2m_y+1} \quad (B10)$$

$\mathcal{H}_{n_i}$ (i=x,y) is the Hermite polynomials with $n_i, \in \mathbb{N}$ are the quantum number. $L_n^\lambda(\varsigma)$ is the associated orthogonal Laguerre polynomials with $m_i \in \mathbb{Z}$. In our work we will correlate the correlation length $L$ and the fluctuation potential $V_0$ with the localization length $R_{cm}$ and the confinement energy $\hbar\omega_{cm}$ respectivelly such as $L = pR_{cm}$, $V_0 = \kappa\hbar\omega_{cm}$ with $p\epsilon\mathbb{R}_+$ and $\kappa\epsilon\mathbb{N}^*$

Note :Here unlike the ref [6] the numbering of the localized excitons state have been done by descending in energy

**Figures**

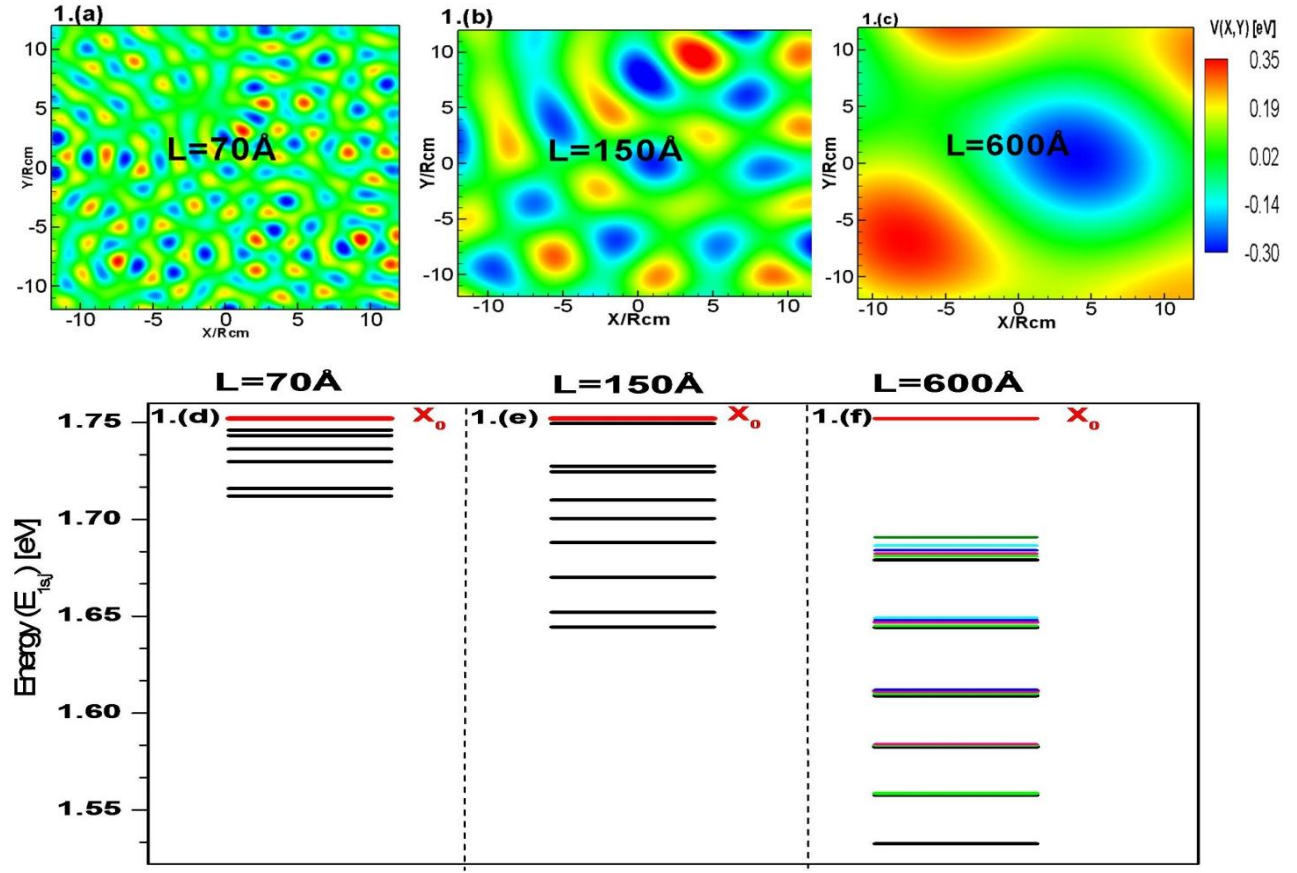

FIG. 1. Spatial map of disorder Potential: (a,b,c )showing lakes-and-hills inhomogeneous landscape, (d,e,f) the corresponding Eigenenergies of the localized excitons obtained by the numerical diagonalization of the matrix resulting from the projection of the Hamiltonian H and localized in the disorder potential given by equation (4). We restrict ourselves only to the states E1s;j smaller than the free exciton energy. The values of $N = 1000$, $V_0 = 110 meV$ , $R_{cm} = 30$ Å are fixed, and we vary le correlation length L. ; a,d) L=70 Å (simulation I ); b,e) L=150 Å (simulation II) and c,f) L = L=600 Å (simulation III).

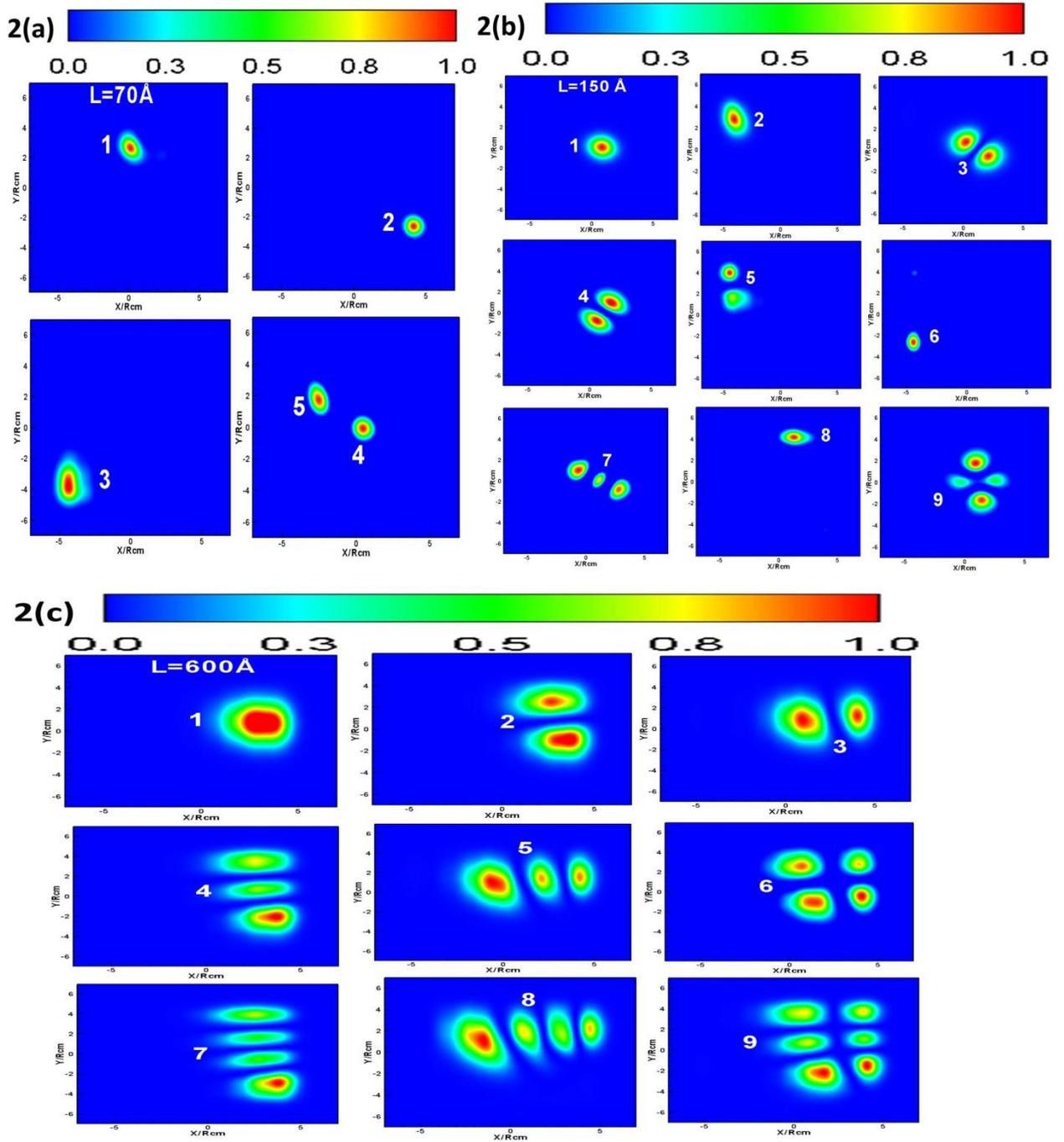

FIG. 2. Probability density distribution of selected excitonic states for fixed values of $N = 1000$, $V_0 = 110 meV$, $R_{cm} = 30$ Å and various values of L ; a) L=70 Å (simulation I ) ; b) L=150 Å (simulation II) and c) L = L=600 Å (simulation III). The color scale is the normalized probability, which show the spatial distribution of the main low-lying states contributing to the LXj .

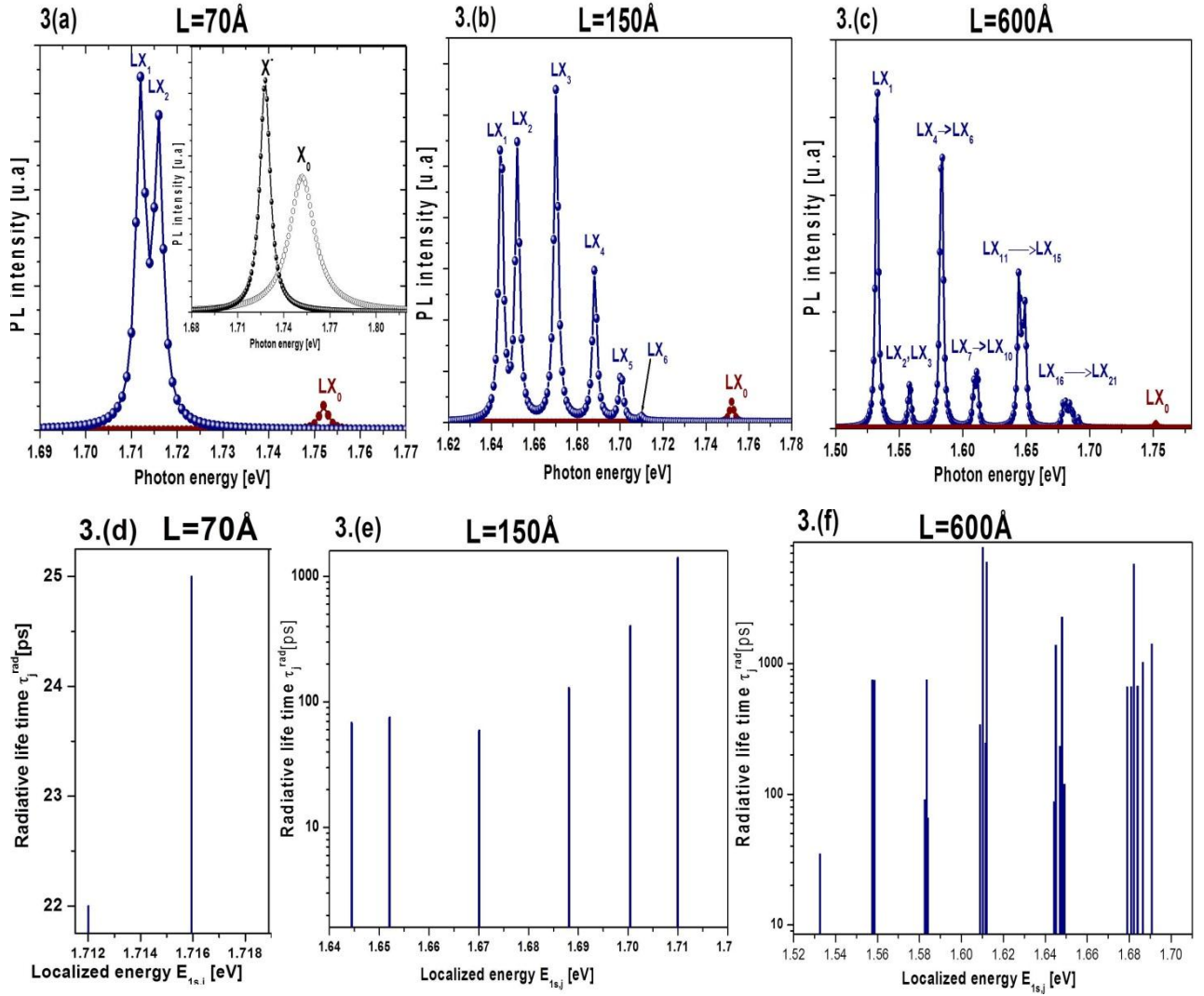

FIG. 3. (a,b,c )PL spectrum of defect from monolayer WSe2 LXj ,(d,e,f) the corresponding radiative lifetime calculating using the equation (15)for fixed values of $N = 1000$, $V_0 = 110 meV$, $R_{cm} = 30$ Å and various values of L ; ; a,d) L=70 Å (simulation I ); b,e) L=150 Å (simulation II) and c,f) L = L=600 Å (simulation III).;in the inset PL spectrum of the neutral free exciton and trion from a pristine region of a monolayer WSe2 calculated in the previous study [59]. The observed low-energy PL peaks are assigned to emissions from localized excitons trapped in the random potential induced by disorder. Note that the localized states are located below the neutral free exciton peak and the values of radiative lifetime are in the range of ten picosocond to nanosocond.

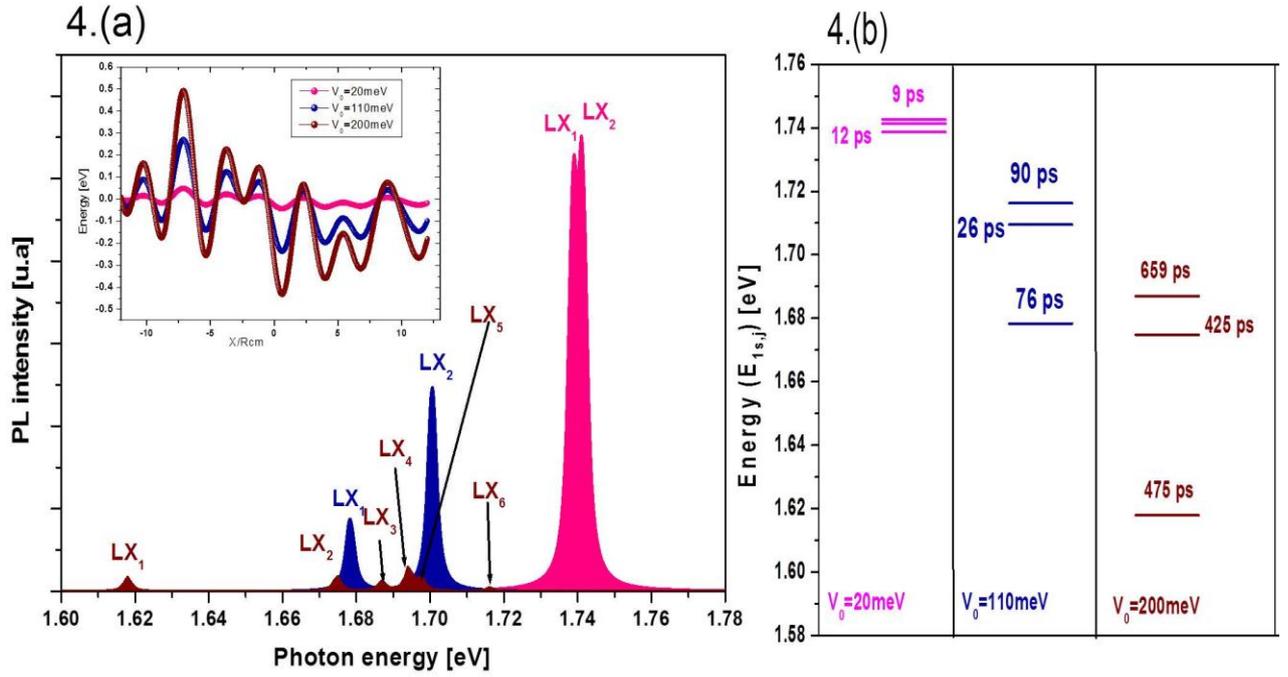

FIG. 4. a) PL spectra of a WSe2 monolayer for different value of $V_0$ (20meV (pink), 110meV (blue), and 200meV (brown)) and for fixed, L = 90 Å, $R_{cm} = 30$ Å and $N = 1000$. b) The energies of the localized exciton obtained by the numerical diagonalization of the matrix resulting from the projection of the Hamiltonian H and the corresponding radiative life time for the three first states.

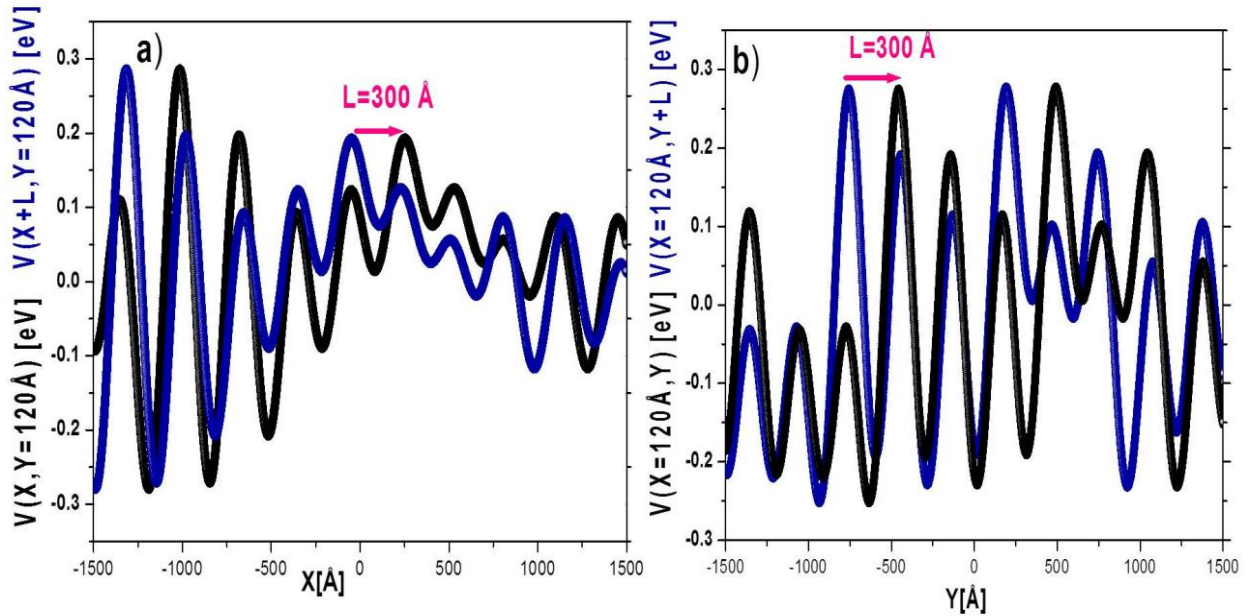

**Fig S1:** a) The variation of the disorder potential V(X, Y=120Å) (black) and V(X+L,Y=120Å) (blue) as a function of center of mass coordinate X and for Y=120Å. b): The variation of the disorder potential V(X=120Å,Y) and V(X=120Å,Y+L) as a function of center of mass coordinate

Y and for X=120Å. The parameters used are $V_0 = 110 meV, N = 500, R_{cm} = 30$Å and L = 300Å.

**TABLE**

TABLE 1. The influence of the dielectric environment on the radiative lifetime of LX1 :of WSe2 monolayer exposed to the air and deposited on a different substrates: the silicon nitride ($\mathcal{E}_{sub}$= 7) ,SiO2 ($\mathcal{E}_{sub}$= 3,9) ,suspended monolayer WSe2(vacuum) ,($\mathcal{E}_{sub}$ = 1) and monolayer encapsulated in hbN at T = 4 K for fixed values of $V_0$ = 110meV , ,L = 90 Å, and $R_{cm}$ = 30 Å .

|  | Encapsulated monolayer | Supported monolayer | | suspended monolayer |
|---|---|---|---|---|
| substrate | hbN | $Si_3N_4$ | SiO2 | vacuum |
| Radiative lifetime | 350ps | 288 ps | 76 ps | 4 ps |